| | |
|---|---|
| Title | PTFE Surface Etching in the Post-discharge of a Scanning RF Plasma Torch: Evidence of Ejected Fluorinated Species |
| Authors | Thierry Dufour, Julie Hubert, Pascal Viville, Corinne Y. Duluard, Simon Desbief, Roberto Lazzaroni, François Reniers |
| Affiliations | [1] Faculté des Sciences, Service de Chimie Analytique et de chimie des Interfaces, Université Libre de Bruxelles, CP-255, Bld du Triomphe, B-1050 Bruxelles, Belgium<br>[2] Service de Chimie des Matériaux Nouveaux (SCMN), Centre de Recherche en Sciences des Matériaux Polymères, Université de Mons-Hainaut, 20 Place du Parc, 7000 Mons, Belgium |
| Ref. | Plasma Processes and Polymers, 2012, Vol. 9, Issue 8, 820-829 |
| DOI | http://dx.doi.org/10.1002/ppap.201100209 |
| Abstract | The texturization of poly(tetrafluoroethylene) (PTFE) surfaces is achieved at atmospheric pressure by using the post-discharge of a radio-frequency plasma torch supplied in helium and oxygen gases. The surface properties are characterized by contact angle measurement, X-ray photoelectron spectroscopy and atomic force microscopy. We show that the plasma treatment increases the surface hydrophobicity (with water contact angles increasing from 115 to 155°) only by modifying the PTFE surface morphology and not the stoichiometry. Measurements of sample mass losses correlated to the ejection of $CF_2$ fragments from the PTFE surface evidenced an etching mechanism at atmospheric pressure. |

# 1. Introduction

Poly(tetrafluoroethylene) (PTFE) is a hydrophobic material, due to a fully fluorinated backbone ($-CF_2-CF_2-$), with surface properties interesting for biological issues.[1] (Super)hydrophobic polymers could also be used as self-cleaning coatings in many applications such as window glass, cement, textiles, etc.[2–5] Previous studies showed that a surface modification of PTFE by plasma treatment in vacuum could lead to an increase in the water contact angle (WCA) and a roughening of the samples.[6,7] This roughness can be associated to the etching[8–11] of the PTFE surface with, in certain cases, the possible incorporation of oxygen-containing polar groups into the surface.[12]

At low pressure (approx. $5.10^{-2}$ Torr), an increase in the hydrophobicity of the surface was obtained by exposing the PTFE surface to neutrals and electrons originating from a resonance frequency (RF) oxygen plasma.[13,14] At atmospheric pressure, the plasma treatments of PTFE most often lead to a decrease on WCA.[15–18] Recently, our group showed that it was possible to obtain very hydrophobic PTFE surfaces by using an atmospheric plasma torch. The carrier gas was argon, and the necessity to add oxygen gas to the plasma was highlighted.[19]

In this study, the PTFE surface modification is achieved at atmospheric pressure by the post-discharge of a scanning RF plasma torch operating with helium and oxygen. The influence of the torch kinematics parameters (torch to sample gap, time of treatment, etc.) and the plasma parameters (RF power, oxygen flow rate and helium flow rate) onto the surface of PTFE are investigated by X-ray photoelectron spectroscopy (XPS), WCA technique and atomic force microscopy (AFM). Measurements of sample mass losses are correlated with de detection of fluorine species onto an aluminum foil close to the PTFE sample.

## 2. Experimental section

### 2.1. The samples

The PTFE samples were supplied by Goodfellow. After having been cut to 15*15mm$^2$ in size, the samples were cleaned first in pure iso-octane and second in pure methanol, before being exposed to the plasma post-discharge.

### 2.2. The Plasma source

The sample surfaces are exposed to the post-discharge of an RF atmospheric plasma torch, the AtomfloTM 400L-Series, from SurfX Technologies.[20] The controller of this plasma source includes a RF generator (27.12 MHz), an auto-tuning matching network and a gas delivery system with two mass-flow controllers to regulate the helium and oxygen gases fuelling the plasma source. Helium (vector gas) and oxygen (reactive gas) are studied for flow rates ranging from 10 to 20 L.min$^{-1}$ and from 0 to 600 mL.min$^{-1}$, respectively. As presented in Figure 1, the resulting gas mixture enters through a tube connected to a rectangular housing (55mm*20mm*80 mm). Inside, the gas is uniformized through two perforated sheets, then flows down around the left and right edges of the upper electrode and passes through a slit in the centre of the lower electrode. A plasma is struck and maintained between these electrodes by applying an RF power to the upper electrode while the lower electrode is grounded. The RF power commonly used is comprised between 60 and 130 W. The geometry of the slit is described as 'linear' due to the ratio of its aperture length (20mm) to its width (0.8 mm). For all the experiments performed in this paper, the plasma torch was used about 20 min after its ignition (time necessary to reach a stable process and especially a constant temperature.

A robotic system is integrated to the plasma torch, enabling the treatment of large samples located downstream. The scanning treatment is achieved with respect to three degrees of freedom corresponding to the three axes of a cartesian coordinate system. In all our experiments, the plasma source was only moved along one direction.

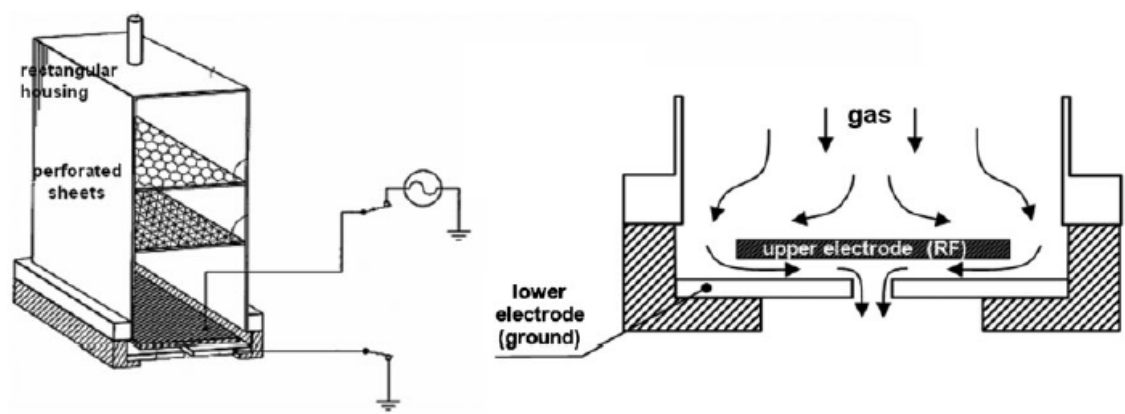

Figure 1. Cross sectional diagram of the RF plasma source (Patent US7329608 Surfx Technologies[19]).

## 2.3. The diagnostics

A drop shape analyzer (Krüss DSA 100) was employed to measure dynamic contact angles of water drops deposited onto PTFE samples. As, in a recent paper,[21] the static method based on the 'Sessile Drop Fitting' was shown to provide information that could not be easily interpreted (this approach was even called 'obsolete'), we measured advancing and receding contact angles by growing and shrinking the size of a single drop on the PTFE surface, from 0 to 15µL at a rate of 30 µL.min$^{-1}$. The WCAs plotted in this paper correspond to the advancing and receding angles.[22] Measurements of receding WCA have been performed on the hysteresis curves by considering the last angle before the deformation of the drop. In all our measurements, a difference of about 5–15° was highlighted between receding and advancing WCA.

To evaluate the chemical composition at the surface of the samples, XPS analyses were performed by using a Physical Electronics PHI-5600 instrument. The base pressure in the analytical chamber was 10$^{-9}$ mbar. Survey scans were used to determine the chemical elements present at the PTFE surface.[23] Narrow-region photoelectron spectra were used for the chemical study of the C 1s, O 1s and F 1s peaks. Spectra were acquired using the Mg anode (1 253.6 eV) operating at 300W. Wide surveys were acquired at 93.9 eV pass-energy, with a five scans accumulation (time/step: 50ms, eV/step: 0.8), and spectra of the C1s peaks at 23.5eV pass-energy with an accumulation of ten scans (time/step: 50ms, eV/step: 0.025). The elemental composition was calculated after removal of a Shirley background line and using the sensitivity coefficients: $S_C$=0.205, $S_F$=1.000 and $S_O$=0.63. The resulting compositions must be taken as indicative and are used only for comparison between the different plasma treatments (with/without O2). They do not reflect the absolute surface composition. In our case, several analyzed regions on the same sample always lead to the same relative compositions.

The surface morphology of the PTFE samples was further analyzed by AFM. This method is very well adapted to characterize the morphology of polymeric surfaces.[24] In this work, all AFM images were recorded in air with a Nanoscope IIIa microscope operated in tapping mode (TM).[25] The probes were commercially available silicon tips with a spring constant of 24–52Nm$^{-1}$, a RF lying in the 264–339 kHz range and a typical radius of curvature in the 5–10nm range. The images presented here are topography signal images recorded with a sampling resolution of 512*512 data points.

The PTFE samples were weighted before and after their plasma treatments to evaluate mass variations. For this, we used the Sartorius BA110S Basic series analytical balance characterized by a 110 g capacity and a 0.1 mg readability. Moreover, during the plasma treatment, every sample was placed on a large aluminium foil. As the aluminium is known to be an efficient fluorine trap,[26] we then analyzed by XPS the presence of fluorinated species on the foil.

# 3. Results and Discussion

Our results are presented and discussed in three sections. The first section introduces results dealing with the kinematic parameters of the plasma torch. We describe how they are likely to influence the efficiency of the surface modification. The second section is dedicated to the plasma parameters, and more specifically to the helium and oxygen flow rates supplying the post-discharge. We discuss their influence by evaluating their impact on WCA measurements and on the PTFE chemical surface composition obtained by XPS. In the last section, we introduce all the results showing that the highlighted modification of the surface can unambiguously be attributed to an etching mechanism.

## 3.1. Influence of the plasma torch parameters

**Kinematic Parameters**

The scanning plasma source moves back and forth along a single axis, alternatively in one direction and in the other. The plasma treatment can therefore be influenced by a lot of kinematic parameters. The first is the scanning length ($L_S$) corresponding to the distance along which the plasma source sweeps the sample. The second is the scanning velocity ($v_S$): the larger is $v_S$, the shorter is the overall process time. The third is the number of scans ($N_S$) and the fourth is the gap ($g_S$) corresponding to the distance between the plasma torch and the upper surface of the PTFE sample. Among those kinematic parameters, one of the most relevant is the number of scans ($N_S$).

In Figure 2 we have plotted the advancing and receding WCA of several samples, each sample being treated for a specific $N_S$. We will only consider the advancing WCA as they are known to be more sensitive to the low-energy components of the surface. The post-discharge was generated with a helium flow rate of 15 L.min$^{-1}$ mixed to an oxygen flow rate of 100 mL min$^{-1}$ and the RF power was 90 W. The applied number of scans ranges from 0 (native PTFE) to 7000. The curve can be divided into three regions, each region representing specific kinetics of the surface modification. The first region ($N_S$=0–30) shows a slight decrease in the advancing WCA from 115° (native surface contact angle) to 102°, thus indicating a slight decay of the surface hydrophobicity. In the second region, from 30 to 500 scans, a strong increase is measured with WCA starting at 102° and levelling up to 136°. Then, for $N_S$=500–7000, a third region is defined in which the WCA still increases but not as strongly as in the second region since the WCA varies from 136 to 154°. Moreover, we note that for $N_S$>4000, the difference between the advancing and receding angles is lower than 5°, corresponding to a small hysteresis that is characteristic of the Cassie-Baxter surface.[27] It indicates a non-wetted contact which was experimentally observed as the drop was unstable and easily slid from the PTFE surface.

Also, we have reported in Figure 2 a second X-axis corresponding to the process time. We can thus estimate that most of the surface modification is achieved after 4 min of plasma treatment. This value of 154° is the maximum WCA obtained with this helium plasma torch. To our knowledge this is the highest contact angle achieved on PTFE using a cold atmospheric plasma treatment. Other atmospheric plasma treatments performed with an RF plasma torch supplied in argon and oxygen show the ability to obtain maximum (static) WCA of 135°.[19] A previous He–$O_2$ atmospheric pressure glow discharge (APGD) allows increasing WCA with a maximum (static) value of 125°.[28] WCA as high as 160° have already been obtained but only for oxygen plasma treatments performed at low pressure (6.67 Pa).[1] To illustrate the existence of this surface modification, six images of deposited water drops are shown in Figure 3, each image corresponding to a specific $N_S$. For instance, the photograph at $N_S$=7000 shows a WCA much higher than those of the first treatments.

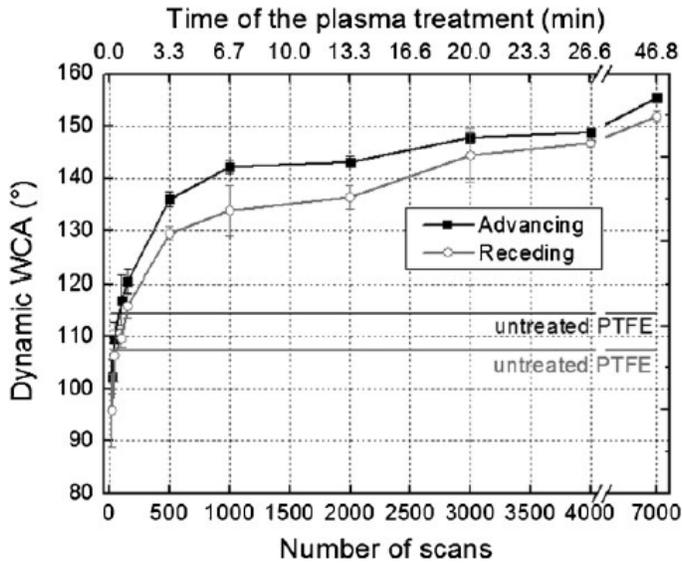

*Figure 2. WCA measured on plasma-treated PTFE samples, versus the number of scans ($L_S$=10mm, $v_S$=25 mm/s, $g_S$=500 mm, F(He)=15 L/min, F($O_2$)=100 mL/min, $P_{RF}$=90 W). The top X-axis gives the corresponding exposure time for the plasma treated samples.*

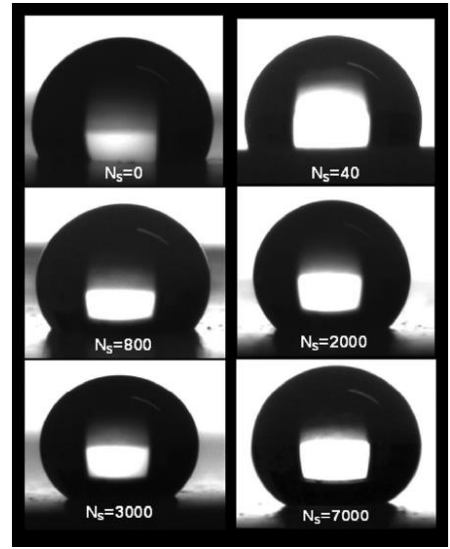

*Figure 3. Droplets of ultra-pure water deposited on different plasma-treated PTFE surfaces; each surface being exposed to a specific number of scans. ($L_S$=10mm, $v_S$=25 mm/s, $g_S$=500 mm, F(He)=15 L/min, F($O_2$)=100 mL/min, $P_{RF}$=90 W).*

The number of scans (or the treatment time) is undoubtedly a relevant kinematic parameter and another one is the gap separating the plasma torch from the PTFE surface. Figure 4 shows the measured WCA versus the gap in the same plasma conditions of the previous treatment, except here a scans number fixed at 1000. This graphic shows that a significant surface modification can only be reached for a gap lower than 3mm. Beyond this distance, between 3 and 6mm, the advancing WCA decreases from 140 to 115°, and for a gap larger than 6mm, the advancing WCA always turns around 115°, therefore indicating the absence of any surface modification.

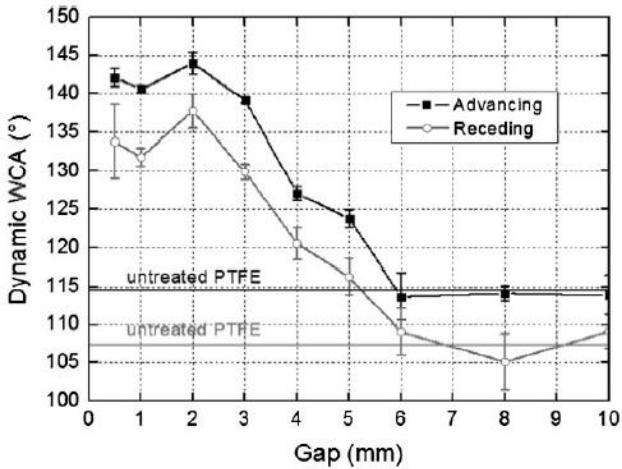

*Figure 4. WCA measured after plasma treatment versus the distance separating the torch from the PTFE surface ($L_S$=10 mm, $v_S$=25 mm/s, $N_S$=1000, F(He)=15 L/min, F($O_2$)=100 mL/min, $P_{RF}$=90 W). The baseline represents the native WCA of PTFE.*

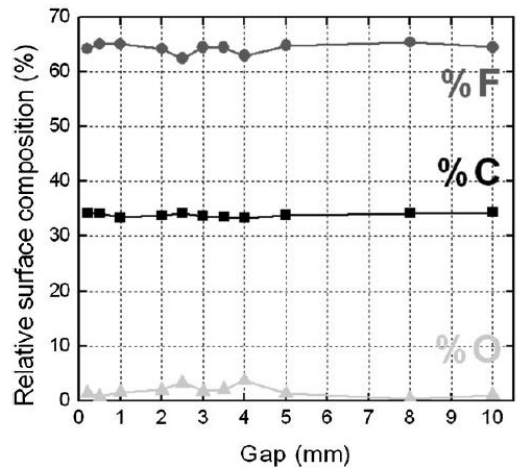

*Figure 5. Surface elementary compositions of He-$O_2$ plasma-treated PTFE samples ($L_S$=10mm, $v_S$=25 mm/s, $N_S$=1000, F(He)=15 L/min, F($O_2$)=100 mL/min, $P_{RF}$=90 W).*

To complete these results, an XPS survey was carried out to determine the relative chemical composition on the PTFE surface. We have plotted in Figure 5 the C 1s, F 1s and O 1s peaks at the respective binding energies: 292.2, 684.9 and 531.6 eV, versus the gap. The distance separating the sample from the plasma torch does not have a significant influence on the surface composition. Indeed, for $g_S$=0.2–10mm, the surface composition is always F=65 at.%, C=34 at.% and O<2 at.%, thus corresponding to an unchanged stoichiometry with a F/C ratio close to 2. The surface stoichiometry was also unchanged in the case of PTFE samples treated with an RF plasma torch fuelled in argon and oxygen.[19] The treatment of an APGD supplied with He-$O_2$ showed a very slight increase in the oxygen concentration (2%) but does not seem to generate any change in surface stoichiometry.[27] High resolution C 1s peaks were also acquired for different gaps. For instance, Figure 6 shows the C 1s peak obtained for $g_S$=500mm and for an untreated PTFE sample. In both cases, the peaks were deconvoluted with the Casa XPS software and the construction of each envelope was always achieved with only two components: $CF_2$– at 292.2 eV and C–C at 284.6 eV. Contrary to PTFE surfaces treated with a low pressure RF plasma supplied in nitrogen,[6] no $CF_3$ component and no evidence of functional groups were detected for the treatment using a helium–oxygen plasma torch. In the case of a low pressure RF plasma supplied in oxygen, small amounts of other fluorinated groups were detected, such as $CF_3$, CF and C–CF at 294, 290 and 287.4 eV, respectively. However XPS analyses were performed with a monochromatic source.[1]

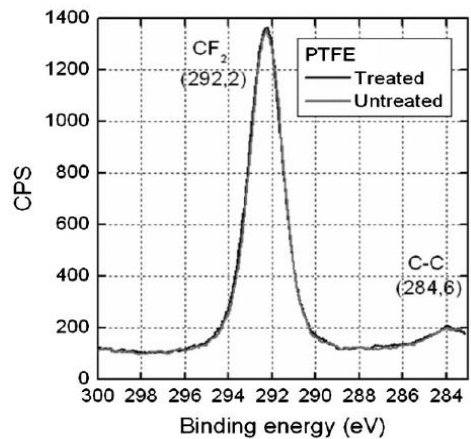

*Figure 6. High resolution C 1s spectrum of He-$O_2$ plasma-treated PTFE for $g_S$=500mm, $L_S$=10 mm, $v_S$=25 mm/s, $N_S$=1000, F(He)=15 L/min, F($O_2$)=100 mL/min, $P_{RF}$=90 W.*

**Gas Flow Rates Parameters**

After having evaluated the essential role of these two relevant kinematic parameters on the PTFE surface modification ($N_S$=1000 and $g_S$=3 mm), we can assess how much the helium and oxygen flow rates modify the PTFE surface. As a first experiment, the helium gas was studied for flow rates ranging from 10 to 20 L.min$^{-1}$ (maximum flow rate of the plasma source). The oxygen flow rate was fixed firstly at 0mL.min$^{-1}$ and then at 100mL.min$^{-1}$ for an RF power of 90 W. The kinematic parameters were the following: $v_S$=25mm.s$^{-1}$, $g_S$=1mm, $L_S$=10mm, $N_S$=1000. The results obtained in these experimental conditions are plotted in Figure 7. When no oxygen is supplied to the plasma torch, the black curve with open squares indicates an average WCA almost 10° lower than the native PTFE WCA (115°) whatever is the helium flow rate. In other words, the helium gas when used alone made the PTFE surface slightly less hydrophobic. The general decrease of WCA by pure helium plasma at atmospheric pressure is in good agreement with literature.[16,18]

By mixing a constant flow rate of oxygen gas (100mL.min$^{-1}$) to the helium gas, the black curve with filled squares of Figure 7 indicates an average WCA clearly higher on the whole helium flow rate range, fluctuating between 140 and 144°. Other studies realized at atmospheric pressure indicated that treatments of PTFE usually lead to a decrease of WCA, as for an $O_2$ atmospheric plasma jet.[15] However, a previous treatment of PTFE by a He-$O_2$ APGD also showed an increase of WCA, but no value higher than 125° was reached.[27] Here again, the carrier gas flow rate does not have a significant impact on the WCA.

As we focus on the hydrophobization process of the PTFE surface, the impact of the oxygen flow rate on the WCA was also studied. A completely different behaviour was obtained when the oxygen flow rate was the variable parameter. We carried out an experiment in which F(O$_2$) was varied from 0 to 200mL.min$^{-1}$ for a constant helium flow rate of 15 L.min$^{-1}$ and an RF power fixed at 90 W. Figure 8 depicts an increase in the WCA with the oxygen flow rate. From 0 to 100mL.min$^{-1}$ of oxygen supplied, the advancing WCA increased strongly from 106 to 142°. Then, for higher oxygen flow rates, a plateau at 142° was reached: any higher oxygen flow rate could no more improve the surface hydrophobicity. Using oxygen as a reactive gas is clearly a necessity to improve the hydrophobicity of the surface. Now, we introduce complementary results corroborating the etching mechanism as being the cause of the surface modification.

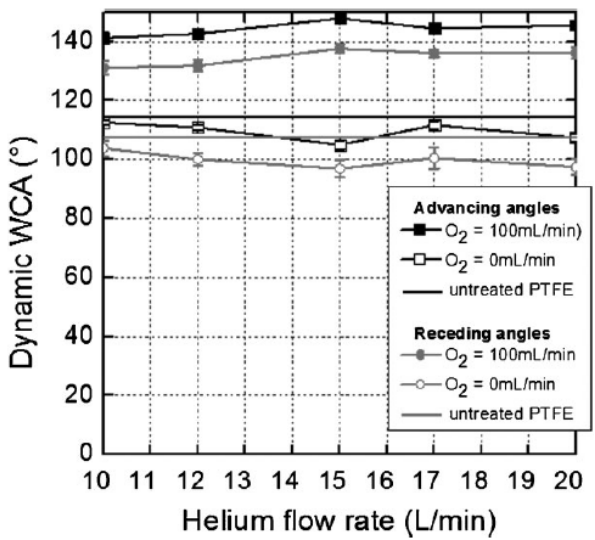
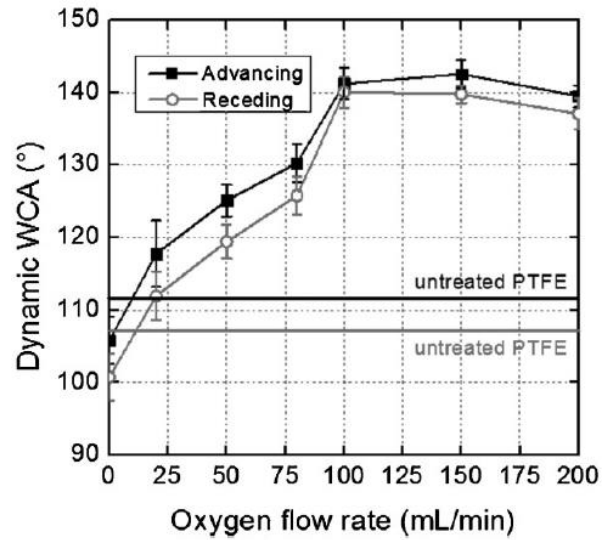

Figure 7. Influence of the helium flow rate on the WCA ($v_S$=25 mm/s, $g_S$=1mm, $L_S$=10mm, $N_S$=1000, $P_{RF}$=90W and F(O$_2$)=0–100 mL/min).

Figure 8. Influence of the oxygen flow rate on the plasma-treated PTFE WCA for $L_S$=10mm, $g_S$=500mm, $N_S$=800, $v_S$=25 mm/s, F(He)=15 L/min, $P_{RF}$=90 W.

## 3.2. Etching of the PTFE surface

In this section we present four complementary experiments highlighting the existence of the surface etching process. The first experiment consists to measure with a microbalance a mass variation of the PTFE samples after their plasma treatment. As all the samples do not have exactly the same dimensions, they do not have exactly the same mass before the plasma treatment. So, to establish a relevant comparison, we need to compare relative mass losses of the samples (RML). This ratio is expressed in ppm and given by the following relation:

$$\text{RML}_{sample} = \left| \frac{\text{Initial mass} - \text{Final mass}}{\text{Initial mass}} \right| \times 10^6$$

As we expected very small mass losses, we cut larger samples (20mm*20mm) compared to the samples of the other experiments (15mm*15 mm) to increase the PTFE areas exposed to the post-discharge. The RML of the sample and its surface WCA are plotted versus the number of scans in Figure 9a and b, respectively. For these two figures, the post-discharge was supplied in helium and oxygen with flow rates of 15 L.min$^{-1}$ and 100 mL.min$^{-1}$, respectively. Several RF powers were tested (from 60 to 150 W). According to Figure 9a, a mass loss is clearly measured after the plasma treatment, mass loss which increases with $N_S$ and the RF power. For instance, at 110W, the RML is 500 ppm for 500 scans and 1500ppm for 3000 scans. Despite the very small losses and the uncertainties attributed to the microbalance accuracy (500 ppm), all the curves clearly show linear trends. The second graphic (Figure 9b) indicates that the WCA strongly increases between 0 and 500 scans and very slowly between 500 and 4000 scans. The maximum WCA obtained is 153° whatever is the RF power supplied. The amount of removed PTFE is therefore proportional to the number of scans but not to the WCA. A maximum value of 150° is reached on the WCA and beyond which the hydrophobicity remains unchanged whatever the RF power and the amount of removed PTFE. A good correlation is achieved between the results of Figure 9 and the results obtained by low pressure (6.67 Pa) plasma treatment of PTFE surfaces in pure oxygen. In this case, the measured mass losses seemed also to increase linearly with the treatment time.[14] Moreover, the evolution of the WCA versus the time treatment is the same as the evolution of the WCA versus the number of scans of our study.

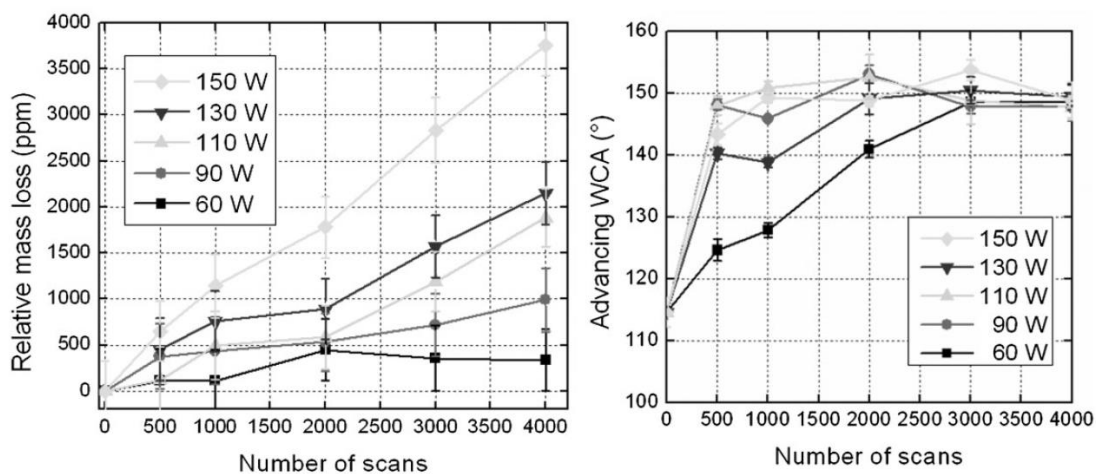

Figure 9. (a) Relative mass losses (ppm) of PTFE samples after plasma treatment and (b) WCA versus $N_S$, for different RF powers. In both cases, the experimental conditions are: $v_S$=25 mm/s, $g_S$=1mm, $L_S$=20mm, F(He)=15 L/min, F($O_2$)=100 mL/min.

As we have measured a mass loss due to the plasma treatment, it should mean (in the case of an etching process) that, at least, volatile fluorinated fragments were removed from PTFE and could be dispersed all around the sample. To verify this assumption, we placed a PTFE sample on a large aluminium foil and exposed it to the scanning post-discharge. The experimental conditions were similar to those previously described in Figure 9. The aluminium foil was then analyzed by XPS to check whether fluorine-containing species could be detected or not, as Al is known to be an efficient fluorine trap.[22] We also analyzed by XPS an aluminium foil unexposed to the post-discharge and an aluminium foil exposed to the post-discharge without any PTFE sample on it. The three corresponding XPS surveys, plotted in Figure 10a, indicate the presence of aluminium, carbon and oxygen but the fluorine species, evidenced by the F 1s peak at 689.4 eV, was only detected on the aluminium foil when it was exposed to the He-$O_2$ postdischarge in presence of a PTFE sample. For more accuracy and a better understanding of this phenomenon, we have performed an XPS mapping, as presented in Figure 10b in which each coloured square represents a fluorine relative concentration at a precise location on the aluminium foil. The white stripped rectangle stands for the PTFE sample during the plasma treatment. The spatial resolution obtained was approximately 1.5*1.5mm$^2$ for each square of the mapping. Figure 10c clearly shows the presence of fluorine species ejected from the PTFE sample with fluorine concentrations of the order of 15 at.% These significant concentrations were detected on the right side (and left side) of the PTFE sample. The green curve in Figure 10c presents the C 1s peak measured 7mm away from the PTFE sample, showing the existence of a $CF_2$ component at 292.2 eV on the aluminium foil.[23] Measuring such a huge amount of volatile PTFE fragments dispersed on an area as large as the sample clearly evidences the etching of the PTFE surface. We have also plotted in Figure 10c, the C 1s peaks for the two previous reference surfaces: a native aluminium foil (black curve) and an aluminium foil exposed to the post-discharge of the plasma torch during 10 min (red curve). In both cases, no $CF_2$ compontent was detected.

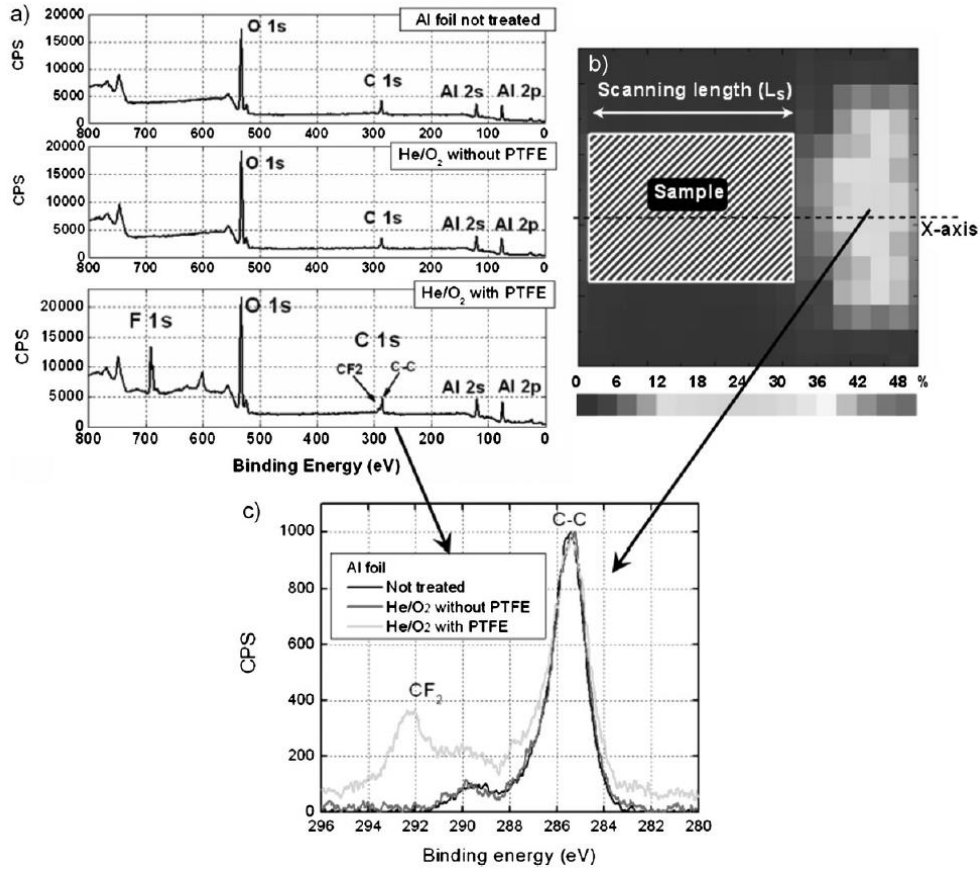

*Figure 10. (a) XPS surveys measured 7mm away from the PTFE sample right boundary. (b) Fluorine relative concentration mapping measured by XPS on the aluminum foil supporting the PTFE sample (white rectangle) after exposure to the scanning plasma source post-discharge. (c) High resolution C 1s spectrum measured 7mm away from the PTFE sample right boundary. The experimental conditions are: $v_S$=25 mm/s, $g_S$=1mm, $L_S$=20 mm, F(He)=15 L/min, F($O_2$)=100 mL/min, $P_{RF}$=90W.*

As the etching would lead to a change in the surface roughness, a consistent method to highlight its existence was to treat a set of nine PTFE samples by plasma for a specific number of scans and, then, to measure their surface roughness by means of AFM in TM. Figure 11 shows seven (5*5mm$^2$) AFM topography (height) images for scans numbers equal to: (a) $N_S$=0 (b) $N_S$=20 (c) $N_S$=150 (d) $N_S$=800 (e) $N_S$=2500 (f) $N_S$=4000 (g) $N_S$=7000, respectively. For images (b) and (c), the PTFE surface exhibits patterns with typical dimensions around 1mm. In the case of images (d) and (e), the previous morphology vanishes to the benefit of smaller and anisotropic structures (alveolar morphology) characterized by average dimensions close to 50 nm. Along with the increase of the number of scans in (f) and (g) images, more alveoli can be distinguished on the surface. Those alveolar patterns are similar to those obtained in the case of PTFE samples treated at low pressure plasma in oxygen.[1] However, at low pressure, the corresponding RMS value is about 500nm (150 nm here) for WCA as high as 160° (150° here).

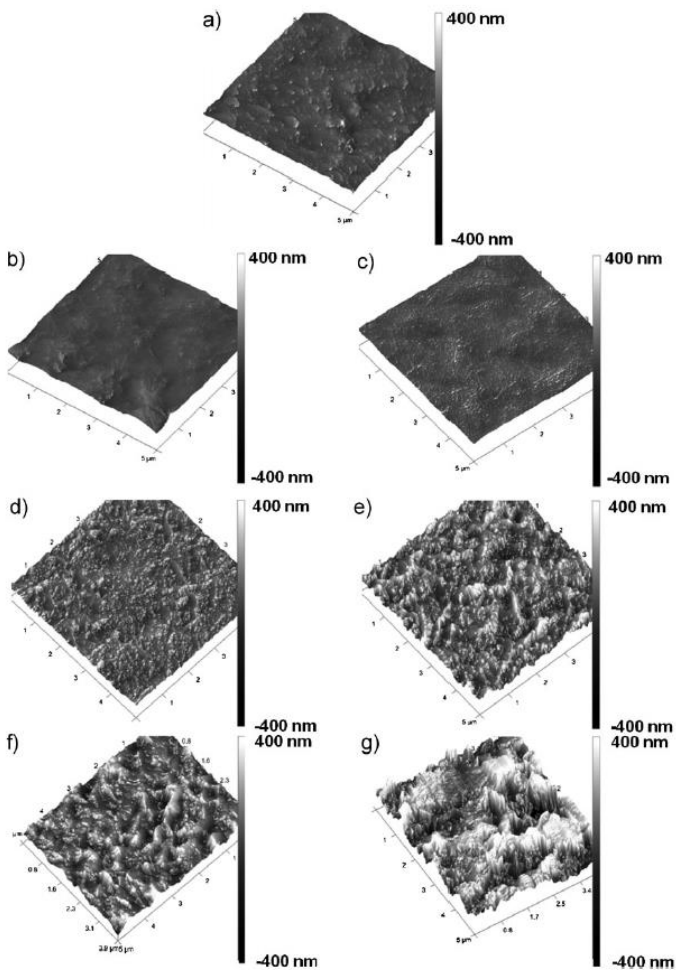

*Figure 11. AFM pictures of plasma-treated PTFE samples for different numbers of scans of the plasma torch: (a) sample not treated, (b) $N_S$=20, (c) $N_S$=150, (d) $N_S$=800, (e) $N_S$=2500, (f) $N_S$=4000 and (g) $N_S$=7000. The experimental conditions are: $v_S$=25 mm/s, $g_S$=1mm, $L_S$=20 mm, F(He)=15 L/min, F(O$_2$)=100 mL/min.*

We have further measured the RMS roughness for each PTFE-treated sample for AFM image sizes of 5*5 mm$^2$. Two regions can be distinguished in Figure 12: for a number of scans comprised between 0 and 150, the RMS slightly decreases from 27 to 21 nm, thus indicating a smoothing of the PTFE surface. This smoothing is in good correlation with the results presented in Figure 2 in which the WCA decreases on the same range ($N_S$=0–150). Then, for numbers of scans comprised between 150 and 7 000, the RMS linearly increases from 21 to 152 nm.

All the previous results are summarized in Table 1 so as to distinguish a three-steps etching process depending on the number of scans. In this table, we only mention the variations of the studied parameters (WCA, mass losses, RMS) by scan. For the first step ($N_S$=0–150), the RMS slightly decreases (–0.04 nm.scan$^{-1}$) whereas the WCA slightly decreases. These results can be interpreted as a cleaning or a smoothing of the surface. The second step ($N_S$=150–500) is in contrast with the first one. Indeed, the WCA strongly increases and so does the RMS (+18.5 nm.scan$^{-1}$) while a constant mass loss of -0.25 ppm.scan$^{-1}$ is evaluated. This step is undoubtedly the most important for the surface hydrophobization, since the resulting WCA reaches already a value as high as 136°. During this step, the mass loss and the increase in the WCA allow us to suggest an anisotropic etching of the surface along the vertical direction, while patterns keep the same lateral characteristic dimensions. The third and last step ($N_S$=500–7000) presents the same mass loss rate and the same RMS rate as previously with however a slower increase in the WCA (only 2.7*10$^{-3}$ °.scan$^{-1}$). Continuing the etching process does not modify the surface hydrophobization significantly because a critical depth of the pattern has almost been reached at the end of step 2, which corresponds to the wetting depth of the Cassie–Baxter model.

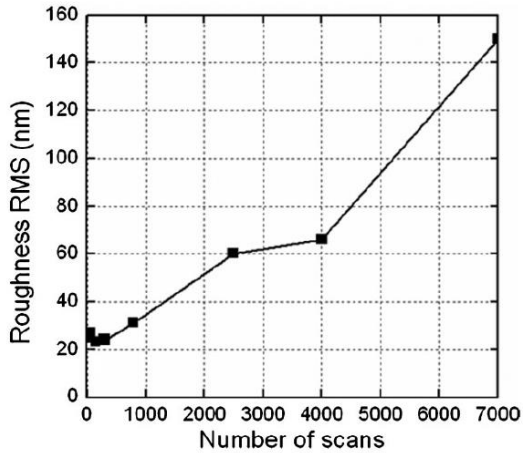

Figure 12. RMS roughness of plasma-treated PTFE samples versus $N_S$. The AFM analysis area was 5*5 mm².

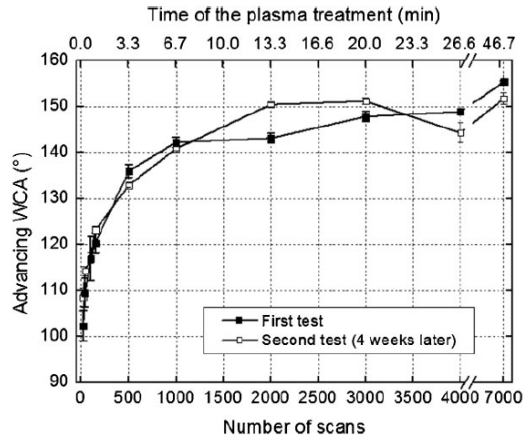

Figure 13. WCA ageing study of PTFE samples treated by a He–O₂ post-discharge (F(He)=15 L/min, F(O₂)=100mL/min, $P_{RF}$=90W, $v_S$=25mm/s, $g_S$=1mm, $L_S$=20mm).

| Parameters | Number of scans | | |
|---|---|---|---|
| | Step 1 | Step 2 | Step 3 |
| | (0–150) | (150–500) | (500–700) |
| WCA (Figure 2) | $-6 \times 10^{-2} \,°\, \text{scan}^{-1}$ | $+4.3 \times 10^{-2} \,°\, \text{scan}^{-1}$ | $+1.8 \times 10^{-3} \,°\, \text{scan}^{-1}$ |
| Mass losses (Figure 9a) | $-0.25$ ppm scan$^{-1}$ | $-0.25$ ppm scan$^{-1}$ | $-0.25$ ppm scan$^{-1}$ |
| RMS (Figure 12a) | $+0.07$ nm scan$^{-1}$ | | $+18.5$ nm scan$^{-1}$ |
| Etched depth (Figure 12b) | $-0.66$ nm scan$^{-1}$ | | $+0.12$ nm scan$^{-1}$ |
| Schemas | | | |

Table 1. Synthesized table of the previous results explaining the etching process.

## 3.3. Ageing study of the PTFE surfaces

The third and last experiment carried out in this work was an ageing study of the sample. We have previously described an experiment consisting of treating samples for different number of scans and then measuring the WCA of their surfaces (see Figure 2). The samples were stored in air during four weeks. Figure 13 represents these last results (red curve with open circles) compared to the first measurements obtained four weeks earlier (black curve with the filled squares). The two experimental curves match, always defining the same three regions delimited on the same previous ranges ($N_S$=0–50, $N_S$=50–500, $N_S$=500–7000). These results have also been corroborated by an XPS analysis: the relative surface compositions of fluorine, carbon and oxygen for an ageing period covering four weeks are presented in Figure 14. In the case of the fluorine and carbon peaks, there is a consistent match between the tests realized four weeks earlier and the reference time. In the case of oxygen, we notice a slight difference, in that a small proportion of oxygen (<2%) is measured by XPS four weeks after the plasma treatment. Therefore, the anisotropic structures previously observed by AFM (see Figure 11) can be considered as steady over time.

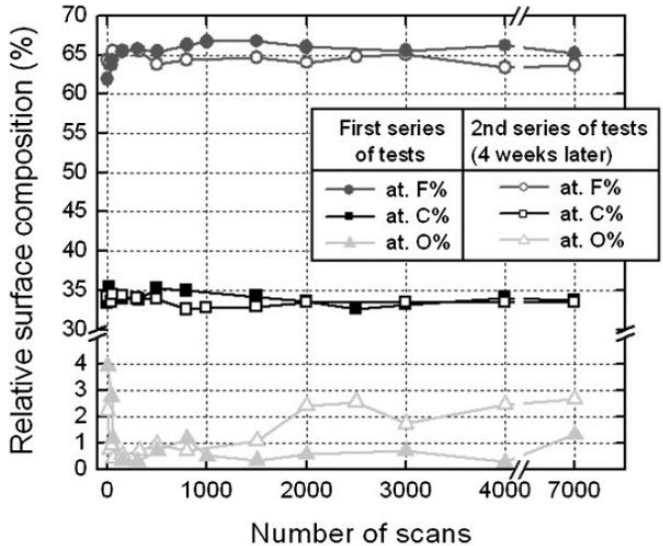

*Figure 14. Relative surface composition of a plasma-treated PTFE sample versus the number of scans of the plasma torch. Every elemental component has been measured twice for an intermediate time lag of 4 weeks ($F(He)$=15 L/min, $F(O_2)$=100mL/min, $P_{RF}$=90W, $v_S$=25mm/s, $g_S$=1mm, $L_S$=20mm).*

## 4. Conclusion

In this paper, we have shown unambiguously that the post-discharge of an atmospheric plasma torch operating with helium and oxygen etches the surface of PTFE, ejecting $CF_2$ fragments, which lead to an increase of the surface roughness, and therefore to an increased hydrophobicity. A good correlation between the sample mass losses and the presence of ejected $CF_2$ fragments close to the sample highlighted the existence of an etching mechanism at atmospheric pressure. The resulting etched surface presents a different morphology but the same composition. Moreover, by increasing the number of scans, we evidenced a significant increase in the WCAs, 154°, which is the maximum angle value obtained in our conditions (i.e. highest number of scans and sample mass losses). According to the literature, this maximum of 154° has never been obtained until now and remains the highest contact angle achieved on PTFE using a cold atmospheric plasma treatment. By increasing the number of scans, the WCAs can be limited while the roughness still increases and the sample mass losses as well. The correlation of these three results indicated that the surface etching was not isotropic as expected but oriented vertically in depth. In outlook, a correlation between the modification of PTFE and the electrical properties of the post-discharge will be attempted.

## 5. Acknowledgements

This work was part of the I.A.P (Interuniversitary Attraction Pole) program financially supported by the Belgian Federal Office for Science Policy (BELSPO). This work was also financially supported by the FNRS (Belgian National Fund for Scientific Research), Région Wallonne (OPTI2MAT Excellence Program) and the European Commission (FEDER – Revêtements Fonctionnels).